\documentstyle[12pt]{article}
\begin{document}
\def\to{\rightarrow}
\def\PRL{Phys.~Rev.~Lett.~}
\def\PRD{Phys.~Rev.~D}
\def\NP{Nucl.~Phys.~}
\def\PL{Phys.~Lett.~}
\def\PLB{Phys.~Lett.~B}

\begin{flushright}
EFI 93-43 \\
TECHNION-PH-93-31\\
hep-ph/9308308 \\
August 1993 \\
\end{flushright}

\bigskip
\medskip
\begin{center}
\large
{\bf Framework for Identification of Neutral ${\bf B}$ Mesons}

\bigskip
\medskip

\normalsize
{\it Michael Gronau} \\

\medskip
{\it Department of Physics, Technion-Israel Institute of Technology \\
Technion City, 32000 Haifa, Israel}

\bigskip
and
\bigskip

{\it Jonathan L. Rosner \\
\medskip

Enrico Fermi Institute and Department of Physics \\
University of Chicago, Chicago, Illinois 60637 } \\

\bigskip
\bigskip
{\bf ABSTRACT}

\end{center}

\begin{quote}

We introduce a method for the study of CP-violating asymmetries in tagged
states of neutral $B$ mesons with arbitrary coherence properties.  A set of
time-dependent measurements is identified which completely specifies the
density matrix of the initial state in a two-component space with basis vectors
$B^0$ and $\overline B^0$, and permits a determination of phases in the
Cabibbo-Kobayashi-Maskawa matrix. For a given tagging configuration, the
measurement of decays both to flavor eigenstates and to CP eigenstates provides
the necessary information.

\end{quote}

\newpage

The search for CP violation in systems involving $B$ mesons is being pursued
using several approaches. Decay modes for which a CP-violating rate asymmetry
would be particularly easy to interpret include those of neutral $B$ mesons to
CP eigenstates such as $J/\psi ~ K_S$ and $\pi^+ \pi^-$. However, one is
required to identify the flavor of the decaying meson: was it a $B^0 ( =
\bar{b} d) $ or $\overline{B}^0 (= b \bar{d})$ at the time of production?

One suggestion for neutral $B$ meson identification involves the reaction
\begin{equation}
e^+ + e^- \to \Upsilon (4S) \to B^0 + \overline{B}^0 ~~~,
\end{equation}
in which the flavor of one meson is identified by correlating it with the
semileptonic decay products of the other. In this configuration, with $C(B
\overline B) = -1$, any CP-violating asymmetry is odd in the difference of
times of the two decays [1,2], so asymmetric storage rings may be needed to
provide the necessary time resolution.  It has also been proposed [2] to
produce the $B \overline B$ pair at a slightly higher energy, in association
with an extra photon from $B^*$ decay, in order to produce a state with $C(B
\overline B) = 1$ to overcome this difficulty.  In both cases the leptonic
``tag'' of a neutral $B$ gives rise to an associated pure $B^0$ or
$\overline{B}^0$.

Another method for determining the flavor of a neutral $B$ relies on the
semileptonic decay of a meson or baryon containing a $b$ quark produced at high
energies in association with the neutral $B$ and with many other particles.
Here the flavor of the decaying $B$ is misidentified part of the time as a
result of $B^0 - \overline{B}^0$ mixing, but in contrast to the case of
production at and just above the $\Upsilon(4S)$ resonance, the initial state of
the decaying neutral $B$ is usually assumed to be an {\it incoherent} mixture
of particle and antiparticle [3].

In order to avoid having to measure the decay of the associated particle, it
has been proposed [4] to study the correlation of the decaying neutral $B$ with
charged pions produced nearby in phase space. This method has been used to
identify neutral $D$ mesons though the decays $D^{*+} \to \pi^+ D^0$ or $D^{*-}
\to \pi^- \overline{D}^0$ [5].  As in semileptonic tagging at high energy, the
initial $B^0$ and $\overline{B}^0$ states were assumed in Ref.~[4] to be
incoherent with one another.

In the present paper we develop a way to identify a neutral $B$ meson which
embraces the above examples but allows one to test experimentally for arbitrary
coherence properties of the initial state.  The method relies upon the
description of the initial tagged state in terms of a density matrix, in a
two-component ``quasispin''
space spanned by $B^0$ and $\overline  B^0$.  The two-component
description is standard for kaons (see, e.g., Refs.~[6]), and has been
applied as well to the case of $B$'s [7].

We have found a set of time-dependent measurements of decays of a tagged $B$
which completely specifies the density matrix, permitting one to obtain phases
in the Cabibbo-Kobayashi-Maskawa (CKM) matrix through a study of CP-violating
decay rate asymmetries.  Each tagging procedure specifies a different density
matrix.  Previously it was necessary to anticipate the properties of such a
matrix with the help of theoretical assumptions. Our result avoids the need for
any such procedure, and provides an answer to the question of
coherence of the initial state.

In brief, we advocate the measurement of $B$ decays both to states
of identifiable flavor (such as $J/\psi K^{*0}$, with $K^{*0} \to K^+ \pi^-$)
and to eigenstates of CP (such as $J/\psi K_S$). The observables
which specify the initial density matrix are:  (1) a
modulation amplitude of the time-dependent term; (2) a shift in phase of the
time-dependent term, and (3) an overall intensity associated with the
production of the CP eigenstate. Once these quantities are measured, one
not only determines the density matrix, but also obtains valuable
information about weak phases.

We now define the basis states and the density matrix.  We choose
\begin{equation}
| B^0 \rangle = \left[ \begin{array}{c} 1 \\ 0 \\ \end{array} \right]~~~,~~
| \overline B^0 \rangle = \left[ \begin{array}{c} 0 \\ 1 \\ \end{array}
\right]~~~.
\end{equation}
A density matrix $\rho$ allows one to discuss incoherent and coherent states in
a unified manner.  The most general such $2 \times 2$ matrix has the form
\begin{equation}
\rho = \frac{1}{2} \left[ 1 +  {\bf Q \cdot \sigma} \right]~~~,
\end{equation}
where ${\bf Q}$ is a vector describing polarization in quasispin space,
satisfying ${\bf Q}^2 \leq 1$, and $\sigma_i~(i = 1,~2,~3)$ are the Pauli
matrices. We give two examples:

(1) Any pure state corresponds to a linear combination of $B^0$ and
$\overline{B}^0$ with arbitrary complex coefficients, whose sum of absolute
squares equals unity.  Such a state can be denoted by a density matrix with
$Q \equiv |{\bf Q}| = 1$.

(2) An arbitrary incoherent combination of $B^0$ and $\overline{B}^0$ with
relative probabilities $P_1$ and $P_2 = 1 - P_1$ corresponds to a diagonal
density matrix with $Q_1 = Q_2 = 0,~Q_3 = 2P_1 - 1$. This is the case we
considered to hold in Ref.~[4].

As a special case of either (1) or (2), one describes the density matrices for
initial $B^0$ and $\overline{B}^0$ by diag(1,0) and diag(0,1), respectively.

The probability for a transition from an initial state denoted by the density
matrix $\rho_i$ to a final state denoted by $\rho_f$ is then
\begin{equation}
I(f) = {\rm Tr}~(\rho_i {T}^{\dag} \rho_f T)~~~,
\end{equation}
where $T$ is the amplitude which time-evolves the state from $i$ to $f$.  Here
we shall take $f$ to denote an arbitrary coherent superposition of $B^0$ and
$\overline{B}^0$ at time $t$.  We shall also be able to write a density matrix
$\rho_f$ which takes account of the {\it decay} of this superposition.

The time-evolution can be visualized by transforming from the $B^0,
\overline{B}^0$ basis to that of mass eigenstates.  We shall neglect
differences between lifetimes of these two eigenstates [1], and denote them by
$B_L$ (``light'') and $B_H$ (``heavy''):
\begin{equation}
|B_L \rangle = p |B^0 \rangle + q |\overline{B}^0 \rangle ~~~,~~
|B_H \rangle = p |B^0 \rangle - q |\overline{B}^0 \rangle ~~~.
\end{equation}
Here $p$ and $q$ are approximately of unit magnitude [1].
We adopt a phase convention in which the $B^0 - \overline{B}^0$ mixing
amplitude is real, so that $ p = q = 1/\sqrt{2}$.  This differs from a more
standard convention [1] by a phase which we shall take into account when
calculating amplitudes for decays of $b$ quarks. The transformation between
flavor eigenstates and mass eigenstates is then implemented by the unitary
matrix
\begin{equation}
U \equiv \frac{1}{\sqrt{2}} \left[ \begin{array}{r r}
1 & 1 \\ 1 & - 1 \\ \end{array} \right]
= \frac{1}{\sqrt{2}} (\sigma_1 + \sigma_3)~~~.
\end{equation}
The matrix describing the time evolution is diagonal in the mass eigenstate
basis:
\begin{equation}
e^{-i M_D t} \equiv e^{- \Gamma t/2}
{\rm diag}(e^{-i m_L t}, e^{-i m_H t})
= e^{- \Gamma t/2} e^{- i \bar m t} e^{i \sigma_3 \Delta m t /2}~~~,
\end{equation}
where $\bar m \equiv (m_H + m_L)/2,~\Delta m \equiv m_H -m_L$.
Thus the time evolution operator $T$ in the $B^0, \overline{B}^0$ basis is
just $T = {U}^{\dag} e^{-i M_D t} U = e^{- \Gamma t/2}
e^{-i \bar m t} e^{i \sigma_1 \Delta m t/2}$.

The trace for the transition probability $I(f)$ can be computed
by applying the matrices $U$ and ${U}^{\dag}$ to the initial
and final density matrices instead of to $e^{-i M_D t}$.  Defining
$\rho' \equiv U \rho {U}^{\dag}$, we find that the effect of $U$ is to rotate
${\bf Q}$ into ${\bf Q}'$, where
\begin{equation}
Q_1' = Q_3~~~,~~Q_2' = - Q_2~~~,~~Q_3' = Q_1 ~~~.
\end{equation}

We already remarked that states which are pure $B^0$ or pure $\overline{B}^0$
correspond to $Q_3 = \pm 1,~ Q_1 = Q_2 = 0$.  Their transformed density
matrices are $\rho'_f = (1/2)(1 \pm \sigma_1)$, respectively, since then $Q_1'
= \pm 1,~ Q_2' = Q_3' = 0$. The transition probability can now be written in
terms of traces as
\begin{equation}
I(f) = {\rm Tr}~ (\rho_i' e^{i M^*_D t} \rho_f' e^{-i M_D t} )~~~.
\label{treq}
\end{equation}

The states $B^0$ and $\overline{B}^0$ at the time of decay $t$ may be
identified by their decays to states of identifiable flavor, e.g., $B^0 \to
J/\psi K^{*0}$, with $K^{*0} \to K^+ \pi^-$.  We assume that a single weak
subprocess contributes to the decay, which is an excellent approximation for
these final states [1].

With the convention in which the mixing amplitudes $p$ and $q$ in the neutral
$B$ mass eigenstates are real, the weak decay amplitudes for $B^0 \to J/\psi
K^{*0}$ and $\overline{B}^0 \to J/\psi \overline{K}^{*0}$ may be denoted $A
e^{-i \beta}$ and $Ae^{i \beta}$, respectively, where $\beta = {\rm Arg}~(-
V_{cb}^* V_{cd}/V_{tb}^* V_{td})$, and $V_{ij}$ are elements of the
Cabibbo-Kobayashi-Maskawa matrix specifying the charge-changing weak couplings
of quarks.  We find

\begin{equation}
I \left( \begin{array}{c} B^0 \\ \overline{B}^0 \\ \end{array} \right) =
\frac{1}{2} |A|^2 e^{-\Gamma t} \left[ 1 \pm (Q_1' \cos \Delta m t - Q_2' \sin
\Delta m t) \right]~~~.
\label{flavor}
\end{equation}

No reference to $Q_3'$ appears.  The sine and cosine terms may be combined into
a cosine of a phase-shifted argument by defining
\begin{equation}
Q_1' \equiv Q_{\perp}' \cos \delta~~~,~~
Q_2' \equiv Q_{\perp}' \sin \delta~~~,
\label{defs}
\end{equation}
so that
\begin{equation}
I \left( \begin{array}{c} B^0 \\ \overline{B}^0 \\ \end{array} \right) =
\frac{1}{2} |A|^2 e^{-\Gamma t} \left[ 1 \pm Q_{\perp}' \cos (\Delta m t +
\delta) \right]~~~.
\label{flavor1}
\end{equation}
Whenever the initial state is any incoherent mixture of $B^0$ and
$\overline{B}^0$ with relative probabilities $P_1$ and $P_2 = 1 - P_1$,
respectively, one simply sets $Q'_\perp = 2 P_1 - 1$ and $\delta = 0$ in the
above expression.

We consider a charge-symmetric production process (such as $\bar p p$ or
$e^+ e^-$ collisions) in which an arbitrary coherent, partially coherent, or
incoherent combination of neutral $B^0$ and $\overline B^0$ is produced, with
an additional particle of specific charge (such as a charged lepton or pion)
bearing some specific kinematic relation to it.  This relation could consist,
for example, of a lepton with specific transverse momentum or range of
transverse momenta, a definite-mass combination $M(B \pi)$ (or range of such
combinations), a specific angle or range of angles between the $B$ and tagging
particle, or a specific rapidity difference (or range thereof) between the two
particles. It is necessary to determine the relation of this process to the one
in which the charge of the tagging particle is the opposite.

Under the phase convention we have chosen for $B^0$ and $\overline{B}^0$, if we
take $CP|B_L \rangle = |B_L \rangle$ and $CP|B_H \rangle = -|B_H \rangle $, the
charge conjugation operation has the phase
\begin{equation}
C |B^0 \rangle = - |\overline{B}^0 \rangle ~~~;~~
C |\overline{B}^0 \rangle = - |B^0 \rangle ~~~.
\end{equation}
Under charge conjugation, the first and second rows and columns of the density
matrix $\rho$ are interchanged, so that $Q_1 \to Q_1,~Q_2 \to -Q_2,~Q_3 \to
-Q_3$. We then find that the effect of charge conjugation is a rotation by
$\pi$ about the $3'$ direction:
\begin{equation}
C:~~~Q'_1 \to -Q'_1,~Q'_2 \to -Q'_2,~Q'_3 \to Q'_3~~~.
\end{equation}
Therefore, the decay rates $\bar I(f)$ for states tagged with antiparticles are
then given in terms of those $I(f)$ for states tagged with particles by
\begin{equation}
\bar I(f;~Q'_1,~Q'_2,~Q'_3) = I(f;~-Q'_1,~-Q'_2,~Q'_3)~~~.
\end{equation}

For a final state identified as a $B^0$ by its decay to $J/\psi K^{*0}$, the
time-dependent asymmetry is
\begin{equation}
A(J/\psi K^{*0}) \equiv \frac{I(J/\psi K^{*0}) - \bar I(J/\psi K^{*0})}
{I(J/\psi K^{*0}) + \bar I(J/\psi K^{*0})}
= Q'_\perp \cos(\Delta m t + \delta)~~~.
\end{equation}
As a consequence of the assumed charge symmetry of the production process, one
has $\bar I(J/\psi K^{*0}) = I(J/\psi \overline{K}^{*0})$ and $\bar I(J/\psi
\overline{K}^{*0}) = I(J/\psi K^{*0})$.

We can then measure both components $Q'_1$ and $Q'_2$, or, equivalently,
$Q'_\perp$ and $\delta$, using decays to flavor eigenstates. The ALEPH
Collaboration [8] has recently measured a time-dependent asymmetry of the above
form, fitting it under the assumption $\delta = 0$.  It should be possible to
set limits on $\delta$ from this measurement. A way of measuring $Q_3'$ will be
described when we come to discuss CP-violating asymmetries.

The above expressions are quite general and not restricted to any specific
tagging method.  The initial density matrix $\rho$ can refer to any method of
preparation, including specific configurations at or just above the
$\Upsilon(4S)$ resonance in $e^+ e^-$ collisions.

A measurement of $Q'_3$ for neutral nonstrange $B$ mesons can be performed by
utilizing their decays to the specific CP eigenstate $J/\psi~K_S$.  In our
phase convention, the amplitudes for $B^0$ and $\overline{B}^0$ to decay into
$J/\psi K_S$ are $A' e^{-i \beta}/\sqrt{2}$ and $-A' e^{i \beta}/\sqrt{2}$.
Here we have taken into account the intrinsic negative CP of the $J/\psi K_S$
state, and neglected the small CP violation in the kaon system. The density
matrix for the final state is then
\begin{equation}
\rho_{J/\psi K_S} = \frac{1}{2} |A'|^2 \left[ e^{-i \beta}~-e^{i \beta}
\right]~ \left[ \begin{array}{c} e^{i \beta} \\ - e^{-i \beta} \end{array}
\right] = \frac{1}{2} |A'|^2 \left[ \begin{array}{c c} 1 & -e^{2 i \beta} \\
-e^{-2 i \beta} & 1 \\ \end{array} \right]~~~.
\end{equation}
For reference, we also note that
\begin{equation}
\rho_{J/\psi K_L} = \frac{1}{2} |A'|^2 \left[ e^{-i \beta}~ e^{i \beta}
\right]~ \left[ \begin{array}{c} e^{i \beta} \\  e^{-i \beta} \end{array}
\right] = \frac{1}{2} |A'|^2 \left[ \begin{array}{c c} 1 & e^{2 i \beta} \\
e^{-2 i \beta} & 1 \\ \end{array} \right]~~~.
\end{equation}

The expressions for the decay rates for states prepared with particle and
antiparticle tags are then
\begin{eqnarray}
I \left( J/\psi \begin{array}{c} K_L \\ K_S \\ \end{array}\right) & = &
\frac{1}{2} |A'|^2 e^{-\Gamma t} \left\{ 1 \pm [Q'_3 \cos 2 \beta + Q'_\perp
\sin 2 \beta~ \sin (\Delta m t + \delta )] \right\}, \nonumber \\
\bar I \left( J/\psi \begin{array}{c}K_L \\ K_S \\ \end{array}\right) & = &
\frac{1}{2} |A'|^2 e^{-\Gamma t} \left\{ 1 \pm [Q'_3 \cos 2 \beta - Q'_\perp
\sin 2 \beta~ \sin (\Delta m t + \delta )] \right\}. \nonumber \\
\end{eqnarray}
As in the case of decays to the flavor eigenstates, the time-dependent term has
a phase shift $\delta$ and a modulation amplitude $Q'_\perp$. It depends upon
$\sin (\Delta m t + \delta)$ rather than $\cos (\Delta m t + \delta)$.

The decay asymmetry for the $J/\psi K_S$ final state is then
\begin{equation}
A(J/\psi K_S) \equiv \frac{I(J/\psi K_S) - \bar I(J/\psi K_S)}
{I(J/\psi K_S) + \bar I(J/\psi K_S)}
= \frac{- Q'_\perp ~\sin 2 \beta ~\sin(\Delta m t + \delta)}
{1 - Q'_3 \cos 2 \beta}~~~.
\end{equation}

The component $Q'_3$ (which appears even in the absence of CP violation)
is a necessary ingredient in the discussion of possible coherence. It is this
component that leads to correlations between $K_S$ and $K_L$ produced in
$\phi$ decay, as discussed in Ref.~[9].  In order to learn its value, we
measure the rate for $J/\psi K_S$ production (summing over particle and
antiparticle tags, so that the time-dependent terms cancel). We compare this
with the corresponding sum of rates (which also has no time-dependence, and is
independent of $Q'_3$) for production of a flavor eigenstate $K^0$.

Since we cannot observe a $K^0$ directly, we must resort to indirect
means.  If the rate of production of charged and neutral $B$'s is equal, as is
expected to be the case for high-energy $e^+ e^-$ collisions [4], one can
compare the rates for $J/\psi K^+$ and $J/\psi K_S$ production, making
use of the fact that the decays of $B$ mesons to $J/\psi K$ involve the quark
subprocess $b \to c \bar c s$, which conserves isospin [10].

A more general method makes use of the expectation that the $K^0/K^+$ ratio
will be the same as the $K^{*0}/K^{*+}$ ratio. The flavor eigenstates
involving $K^{*0},~K^+$, and $K^{*+}$ are all observable.  Thus we can
determine the ratio $|A'/A|^2$.

In either method we learn the relative normalization of rates for decays to
flavor eigenstates and CP eigenstates.  We thereby determine the magnitude of
the term $Q_3' \cos 2 \beta$, and then use the asymmetry in $J/\psi K_S$ decay
to measure $ \sin 2 \beta$.  With the possibility of a discrete ambiguity
(unlikely for known ranges of CKM parameters), we then obtain $\cos 2 \beta$,
thereby finding $Q'_3$ itself.

The corresponding decay asymmetry for the $\pi^+ \pi^-$ final state is easily
calculated by our method.  We neglect penguin effects [11], which can be dealt
with by studying the $2 \pi^0$ final state.  With our phase convention for $b$
quarks, the result can be obtained by the substitution $\beta \to - \alpha$ in
the corresponding result for the $J/\psi K_L$ final state, where $\alpha = {\rm
Arg}~(-V_{tb}^* V_{td}/V_{ub}^* V_{ud})$. We find
\begin{equation}
A(\pi^+ \pi^-) \equiv \frac{I(\pi^+ \pi^-) - \bar I(\pi^+ \pi^-)}
{I(\pi^+ \pi^-) + \bar I(\pi^+ \pi^-)}
= \frac{- Q'_\perp ~\sin 2 \alpha ~\sin(\Delta m t + \delta)}
{1 + Q'_3 \cos 2 \alpha}~~~.
\end{equation}
Since we have already measured all components of ${\bf Q}'$ and the phase
$\delta$, this result can be used to extract $\alpha$.

To sum up, we have developed a density-matrix formalism for the discussion of
decays of neutral $B$ mesons which have been ``tagged'' in any arbitrary
fashion.  The density matrix is specified in terms of a polarization vector
${\bf Q}$ in the basis labelled by $B^0$ and $\overline  B^0$, or ${\bf Q}'$ in
the basis of mass eigenstates.  It is the components of this polarization
vector which determine all observable CP-violating asymmetries, and which can
be learned by measurements on the time-dependence of decays to $J/\psi K^*$ and
$J/\psi K$. One learns as well the angle $\beta$. Decays to final states such
as $\pi^+ \pi^-$ can then provide information on other phases of the
Cabibbo-Kobayashi-Maskawa matrix.

We are grateful to B.~Kayser, H.~Lipkin, A.~I.~Sanda, and S.~Stone for helpful
discussions, and to D.~G.~Cassel and H.~Tye for extending the hospitality of
the Newman Laboratory of Nuclear Science, Cornell University, during part of
this investigation.  This work was supported in part by the United States --
Israel Binational Science Foundation under Research Grant Agreement 90-00483/2,
by the Fund for Promotion of Research at the Technion, and by the
U.~S.~Department of Energy under Grant No. DE FG02 90ER-40560.

\newpage
\centerline{\bf REFERENCES}

\begin{enumerate}

\item[{[1]}] I.~Dunietz and J.~L.~Rosner, \PRD {\bf 34}, 1404 (1986);
I.~Dunietz, Ann. Phys. (N.Y.) {\bf 184}, 350 (1988);
I.~Bigi and A.~I.~Sanda, \NP {\bf B281}, 41 (1987);
Y.~Nir and H.~Quinn, Ann.~Rev.~Nucl.~Part.~Sci.~{\bf 42}, 211 (1992).

\item[{[2]}] {\it Updated Proposal for a B Factory Upgrade for the Cornell
Electron Storage Ring}, Cornell University report CNLS-CESR-B-UPD-PROP, May,
1993 (unpublished);  {\it Progress Report on Physics and Detector at KEK
Asymmetric $B$ Factory}, KEK Report 92-3, May 1992 (unpublished); {\it An
Asymmetric $B$ Factory Based on PEP}, Lawrence Berkeley Laboratory LBL
PUB-5303, February, 1991 (unpublished).

\item[{[3]}] See, e.g., {\it B Physics at Hadron Accelerators}, Proceedings of
a workshop at Fermilab, Nov. 16--17, 1992, edited by J.~A.~Appel
(Fermilab, Batavia, IL, 1992);  {\it Beauty 93} (Proceedings of the First
International Workshop on $B$ Physics at Hadron Machines, Liblice Castle,
Melnik, Czech Republic, Jan.~18--22, 1993), edited by P.~E.~Schlein,
Nucl.~Instrum.~Meth. {\bf 331}, No.~1 (1993); M. Chaichian and A. Fridman, \PLB
{\bf 298}, 218 (1993).

\item[{[4]}] M.~Gronau, A.~Nippe, and J.~L.~Rosner, \PRD {\bf 47}, 1988
(1993).

\item[{[5]}] S.~Stone, in {\em Heavy Flavors}, edited by A.~J.~Buras and
M.~Lindner (World Scientific, Singapore, 1992).

\item[{[6]}] J.~S.~Bell and J.~Steinberger, in {\it Proceedings of the
Oxford International Conference on Elementary Particles}, 19--25 September,
1965, edited by T.~R.~Walsh {\it et al.} (Rutherford High Energy Laboratory,
Chilton, England, 1966), p.~195;
T.~D.~Lee and C.~S.~Wu, Ann.~Rev.~Nucl.~Sci. {\bf 16}, 511 (1966).

\item[{[7]}] H.~J.~Lipkin, \PLB {\bf 219}, 474 (1989); Weizmann Institute
preprint WIS-93-12, 1993 (unpublished).

\item[{[8]}] ALEPH Collaboration, CERN report CERN-PPE/93-99, June, 1993,
submitted to Phys.~Lett.~B.

\item[{[9]}] H.~J.~Lipkin, Phys.~Rev.~{\bf 176}, 1715 (1968).

\item[{[10]}] H.~J.~Lipkin and A.~I.~Sanda, \PLB {\bf 201}, 541 (1988).

\item[{[11]}] M.~Gronau, \PRL {\bf 63}, 1459 (1989); \PLB{\bf 300}, 163
(1993); M.~Gronau and D.~London, \PRL {\bf 65}, 3381 (1990).
\end{enumerate}
\end{document}